\documentclass[superscriptaddress,12pt,preprint]{revtex4-1}

\usepackage{graphics}
\usepackage{graphicx}
\usepackage{amsmath}
\usepackage{amssymb}

\newcommand{\citeasnoun}[1]{Ref.~\citenum{#1}}
\newcommand{\secref}[1]{Sec.~\ref{sec:#1}}

\newcommand{\figref}[1]{Fig.~\ref{fig:#1}}
\newcommand{\Figref}[1]{Figure~\ref{fig:#1}}
\newcommand{\figreftwo}[2]{Figs.~\ref{fig:#1}--\ref{fig:#2}}
\renewcommand{\eqref}[1]{Eq.~(\ref{eq:#1})}
\newcommand{\Eqref}[1]{Equation~(\ref{eq:#1})}

\usepackage{xcolor}

\begin{document}

\title{Factorized Machine Learning for Performance Modeling of Massively Parallel Heterogeneous Physical Simulations}

\author{Ardavan~Oskooi}
\email{oskooi@simpetus.com}
\affiliation{Simpetus LLC, San Francisco, CA}
\author{Christopher~Hogan}
\affiliation{Simpetus LLC, San Francisco, CA}
\author{Alec~M.~Hammond}
\affiliation{Simpetus LLC, San Francisco, CA}
\affiliation{Georgia Institute of Technology, Atlanta, GA}
\author{M.~T.~Homer~Reid}
\affiliation{Simpetus LLC, San Francisco, CA}
\author{Steven~G.~Johnson}
\affiliation{Simpetus LLC, San Francisco, CA}
\affiliation{Massachusetts Institute of Technology, Cambridge, MA}

\begin{abstract}
  We demonstrate neural-network runtime prediction for complex, many-parameter, massively parallel, heterogeneous-physics simulations running on cloud-based MPI clusters. Because individual simulations are so expensive, it is crucial to train the network on a limited dataset despite the potentially large input space of the physics at each point in the spatial domain. We achieve this using a two-part strategy. First, we perform data-driven static load balancing using regression coefficients extracted from small simulations, which both improves parallel performance and reduces the dependency of the runtime on the precise spatial layout of the heterogeneous physics. Second, we divide the execution time of these load-balanced simulations into computation and communication, factoring crude asymptotic scalings out of each term, and training neural nets for the remaining factor coefficients. This strategy is implemented for \emph{Meep}, a popular and complex open-source electrodynamics simulation package, and are validated for heterogeneous simulations drawn from published engineering models.
\end{abstract}

\maketitle

\section{Introduction}
\label{sec:intro}

\begin{figure}[t]
{\centering \includegraphics[width=1.0\columnwidth]{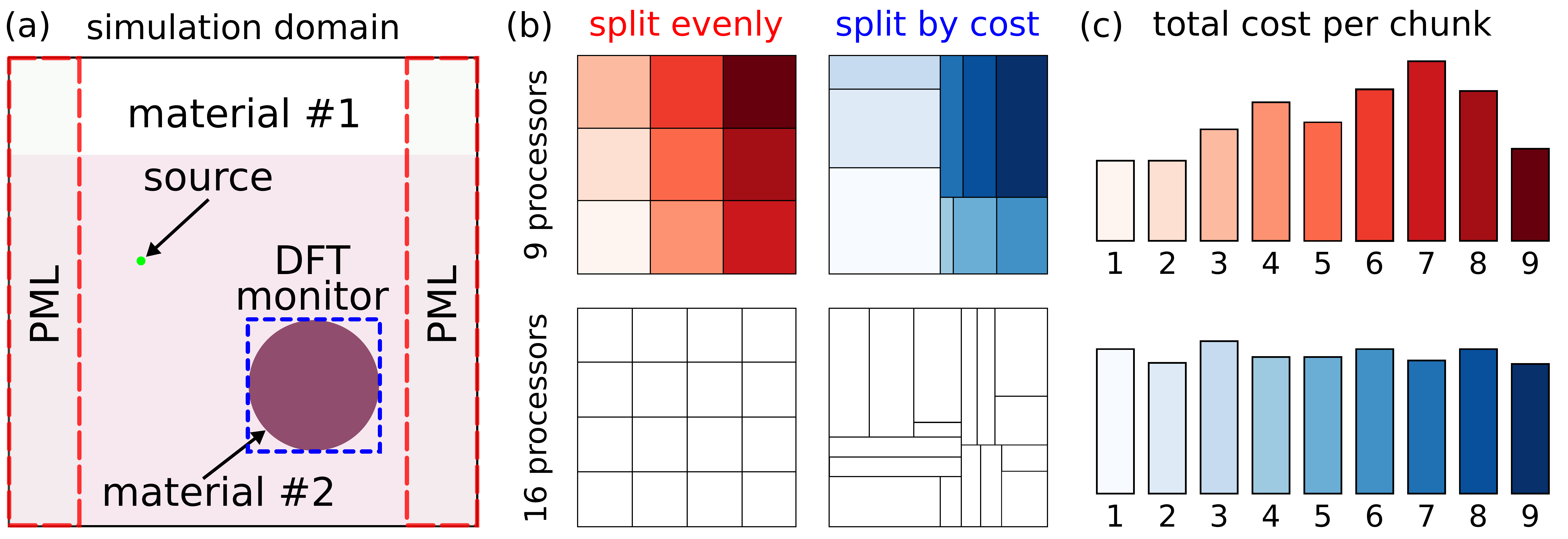}
\par}
\caption{(a) Schematic electrodynamics simulation of materials and calculations that are heterogeneous across the simulation domain (described in \secref{FDTD}), which greatly complicates performance analyses.  (The heterogeneity varies from one simulation to the next.)  This is parallelized by (b) partitioning the domain into one ``chunk'' per processor.  An equal partition (red) results in (c) very unequal computational costs on the different processors.  A data-driven cost estimate (\secref{LB}) results in an unequal-area partition (blue) with load-balanced costs (c, blue). Load-balancing simplifies runtime prediction because the balanced runtime depends mostly on the types of computations performed but not on their spatial distribution.
\label{fig:LB}}
\end{figure}

\begin{figure}[t]
{\centering \includegraphics[width=1.0\columnwidth]{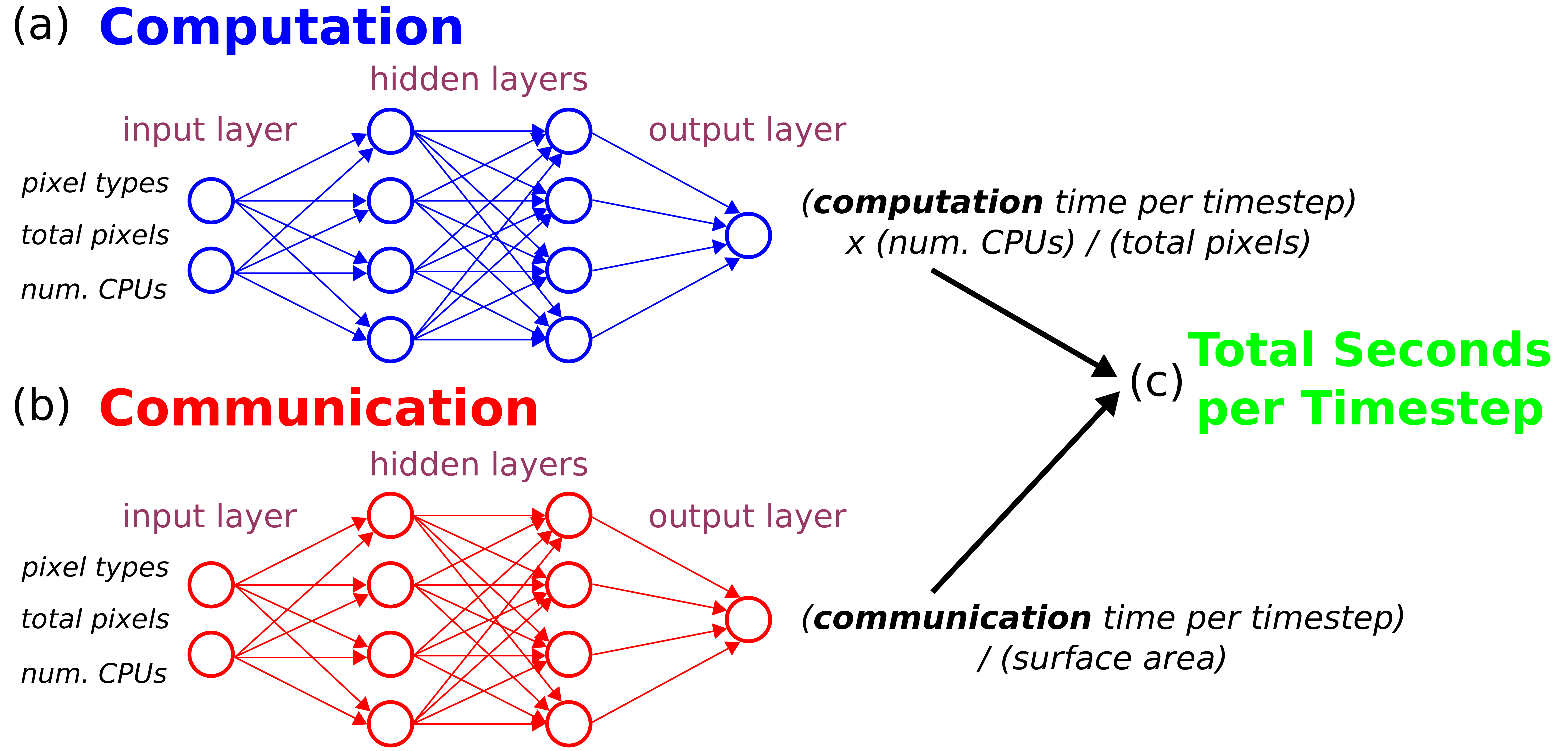}
\par}
\caption{To train a performance predictor with minimal simulation data, we (c) factor the total execution time (per simulation timestep) into computation and communication, and factor out the asymptotic scalings of load-balanced simulations with the numbers of processors (CPUs), pixels (grid points), and surface area [see also \eqref{factorization}].  The remaining (a) computation and (b) communication coefficients are fit to neural networks (\secref{NN}). Since these coefficients represent \emph{corrections} to crude scaling laws, they are much closer to \emph{constants} than raw timings, and hence are easier for a NN to learn.
\label{fig:NN}}
\end{figure}

\noindent For large-scale parallel physics simulations in science and engineering, it is crucial to have a rough estimate of the execution time \emph{in advance} for a given compute platform, in order to allocate system resources and choose simulation parameters efficiently. Ideally, this estimation should be \emph{automated}, but we do not have the luxury of big datasets available in other areas of machine learning: collecting empirical data for such an estimate involves massively parallel computations, which must be re-run for each new hardware platform, so one would like to construct an accurate estimate from as little data as possible. ``Analytical'' performance estimates from detailed hardware models and fine-grained code instrumentation are possible in controlled settings, but it is difficult to apply that approach to large codebases with many features undergoing continual development. In this paper, we present a technique for performance prediction of realistic physical simulations (FDTD electrodynamics in heterogeneous materials, \figref{LB} and \secref{FDTD}), using neural networks (NNs) applied to a carefully \emph{factored} form of the execution time (\figref{NN}) that allows us to train the network using a relatively small number of execution measurements by exploiting crude a priori knowledge of the scaling characteristics of load-balanced simulations (\secref{NN}). Whereas previous works on performance prediction using NNs and other methods (below) often attacked simple model problems in which identical computations occur everywhere in a uniform computational mesh or dataset, divided equally among processors, we address a more realistic practical situation of \emph{heterogeneous physics}: different phenomena or calculations are modeled at different points in the computational domain, as sketched in \figref{LB}. Different materials or data processing will often require vastly disparate computational resources, and merely dividing such a domain into equal-volume chunks for each processor (\figref{LB}b red) can result in an imbalanced computational load (\figref{LB}c red), so that some processors are idle while others complete their work. This both degrades performance and makes performance prediction more difficult since it depends on the precise spatial layout. Hence, we also apply a data-driven approach to \emph{load balancing} (\secref{LB}), in which a small number of simulations are used to estimate the costs of different model components, leading to a new partitioning algorithm that produces unequal domains as needed (\figref{LB}b blue) with nearly equal costs per process (\figref{LB}c blue). This heterogeneity is also an input to the NN (\secref{NN}), and despite the complexity of such unequal-chunk parallel computations we are able to predict the execution time of simulations drawn from real applications with a mean error of around $20\pm 10$\% on Amazon EC2 cloud-computing clusters. Load balancing allows the NN to predict execution based on what \emph{kinds} of physics are present but without needing to know the exact spatial distribution, enabling us to train a 6-input NN with $\sim 10^4$ simulations. We achieve this for a popular free/open-source physics package~\cite{oskooi2010meep} with a complicated C++ codebase (200,000+ lines) and a huge Python-scripted feature set, using minimal code modifications, making us optimistic that similar techniques could be applied to other practical simulation software.

There is a rich literature of previous work on predicting execution times of scientific computations.  As reviewed in \citeasnoun{Sanjay08}, considerable effort has been expended in developing ``analytical'' models that predict code performance based on low-level hardware details (e.g. memory bandwidth etc.) combined with fine-grained (often compiler-assisted) software instrumentation. Such analyses are challenging to apply to large, evolving codebases.  Many authors have also proposed automated performance analyses, either based on polynomial fits \cite{Barnes08,Sanjay08,Lobachev13,Wolf14,Grebhahn16} or NNs \cite{Ipek05,Lee07,Singh07,Balaprakash14,Bernst15,Hieu2016,Nadeem17,Wyatt2018}. However, none of this work considered spatial heterogeneity, and the simulations were typically characterized only by the size of the grid (amount of data) and the number of processors. Some authors considered homogeneous calculations across heterogeneous computing hardware \cite{Bernst15,Nadeem17}, which effectively adds one or two additional inputs to the NN.

Massively parallel scientific simulations necessarily provide only a limited amount of training data.  As the number $n$ of inputs (of the simulation or the hardware) is increased, a larger and larger set of training data is required for a NN to fully characterize the problem space, but acquiring training data is costly in our case: each data point is a large-scale parallel simulation. In order to reduce the amount of training runs required to obtain accurate predictions for heterogeneity with many inputs $\vec{p}\in\mathbb{R}^n$, we \emph{factorize} the execution time $T(\vec{p})$ to exploit crude {\it a priori} knowledge, as depicted in \figref{NN}.   First, we separate $T$ into a sum of computation and communication time, similar to \citeasnoun{Sanjay08} and \citeasnoun{Bernst15}, which requires little or no code instrumentation because in practice all communications pass through a single library such as MPI \cite{MPISpec} that typically provides profiling data already.  The ability to predict computation and communication time separately is useful in its own right, because the computation/communications ratio is a common measure of parallelization efficiency, but it also allows us to apply knowledge of crude scaling laws.   Although a precise formula for the performance of complex software is difficult to obtain, one almost always knows the complexity in an \emph{asymptotic} sense, e.g. whether it scales as $\Theta(N)$ vs.~$\Theta(N^2)$ in the number $N$ of grid points.  By factoring out this asymptotic scaling from the computation and communication terms and only using NNs for the \emph{coefficients} of the scaling, we can train the NNs with much less data: the closer a function is to a constant, the easier it is to learn.  In particular, we factor the time $T$ per FDTD timestep (see \secref{FDTD}) as:
\begin{equation}
T(\vec{p}) = \underbrace{\frac{W(\vec{p}) \times N}{P}}_\mathrm{computation} + \underbrace{C(\vec{p}) \times S}_\mathrm{communication}
\label{eq:factorization}
\end{equation}
where $N$ is the total number of grid points, $P$ is the number of processors, and $S$ is the \emph{maximum} surface area of any processor's subdomain.  (Since timestepping is synchronous, $C\times S$ is determined by the process that requires the \emph{most} communication.)  \Eqref{factorization} factors out the knowledge that the computational work of FDTD scales asymptotically as $\Theta(N/P)$, the number of grid points per processor, and that the communication cost scales roughly with the area.  Of course, these scalings are not exact, so we train NNs for the coefficient functions $W(\vec{p})$ (computation) and $C(\vec{p})$ (communication) as described in \secref{NN}, in order to account for all of the complications that such crude scalings omit.  Moreover, since we are interested in minimizing \emph{relative} error, we actually fit NNs to $\log W$ and $\log C$, as was similarly suggested in \citeasnoun{Barnes08}, and we similarly take logarithms of most inputs.  As described in \secref{NN}, this factorization results in a dramatic improvement in accuracy compared to naively fitting $T$ directly, and many further refinements are possible as discussed in \secref{conclusions}.

\section{Heterogeneous FDTD Electrodynamics Simulations}
\label{sec:FDTD}

\noindent The physical problem that we consider in this paper is a finite-difference time-domain (FDTD) simulation of electromagnetism (EM), as reviewed in \citeasnoun{taflove13} and implemented in the free/open-source package \href{https://meep.readthedocs.io/en/latest/}{\emph{Meep}} \cite{oskooi2010meep}. EM modeling is central to much of science and engineering, from scales ranging from radar (meters) to X-rays (nanometers), for a vast array of applications from photovoltaics to imaging, and the FDTD method is popular because of its generality, flexibility, and scalability to large parallel computations. For a given computational domain (spatial region), it discretizes space into a grid of unknowns (EM fields); on each discretized ``timestep'' the fields everywhere in the grid are updated (via nearest-neighbor interactions) to a new value. A typical simulation problem, e.g. to determine how light waves are absorbed by an image sensor, requires many thousands of timesteps to model the wave propagating all the way through the domain. Moreover, a typical FDTD problem is \emph{heterogeneous}, as depicted in \figref{LB}: different points in the spatial grid require different computations depending on the materials being modeled at each point. The simplest material is vacuum, whereas more complicated updates are required at grid points with nonlinear materials or materials with frequency-dependent responses, for example. Certain points in space might have ``source'' terms generating waves (like a physical antenna). Adjacent to the edges of the domain there are often artificial absorbing layers called PMLs (perfectly matched layers) to inhibit unwanted reflections from the boundaries, which require additional unknowns and more complicated update equations~\cite{Oskooi2011}. Additionally, in some regions one may perform expensive post-processing of the data at every timestep for output analysis, such as accumulating discrete-time Fourier transforms (DFTs) at certain points to obtain scattering spectra~\cite{oskooi2010meep}. This heterogeneity complicates the modeling of execution time---it is not simply a function of the total number of grid points per processor as assumed in many previous works (below)---and it also complicates parallelization.

As mentioned above and indicated schematically in \figref{NN}, one should not simply divide the domain into equal-volume pieces. Instead, we develop a data-driven heuristic cost function for each type of physics/analysis and partition the domain in order to equalize this cost, as explained in \secref{LB}. Many previous works on mesh-partitioning methods minimized only the surface area of the divisions \cite{Devine06} (assuming the cost per grid point is constant), since communications are required between adjacent grid points \cite{oskooi2010meep}; we incorporate this communications scaling in our performance model of \eqref{factorization} and \secref{NN}.  Instead, we employ a generalization of a recursive bisection strategy proposed by \citeasnoun{Berger87} for mesh partitioning with heterogeneous computational loads.  In order to estimate the relative loads, we use a regression technique described in \secref{LB}. Load balancing also simplifies the task of predicting execution time: for an unbalanced simulation, the time will be determined by the slowest process, and predicting this would require the NN to understand the exact spatial distribution of computation costs. In contrast, for a balanced simulation the runtime is determined by the types of computations but not where they occur, allowing a simpler NN (with fewer inputs and training data) to yield accurate predictions.

\section{Data-driven Load Balancing}
\label{sec:LB}

\begin{figure}[t]
{\centering \includegraphics[width=1.0\columnwidth]{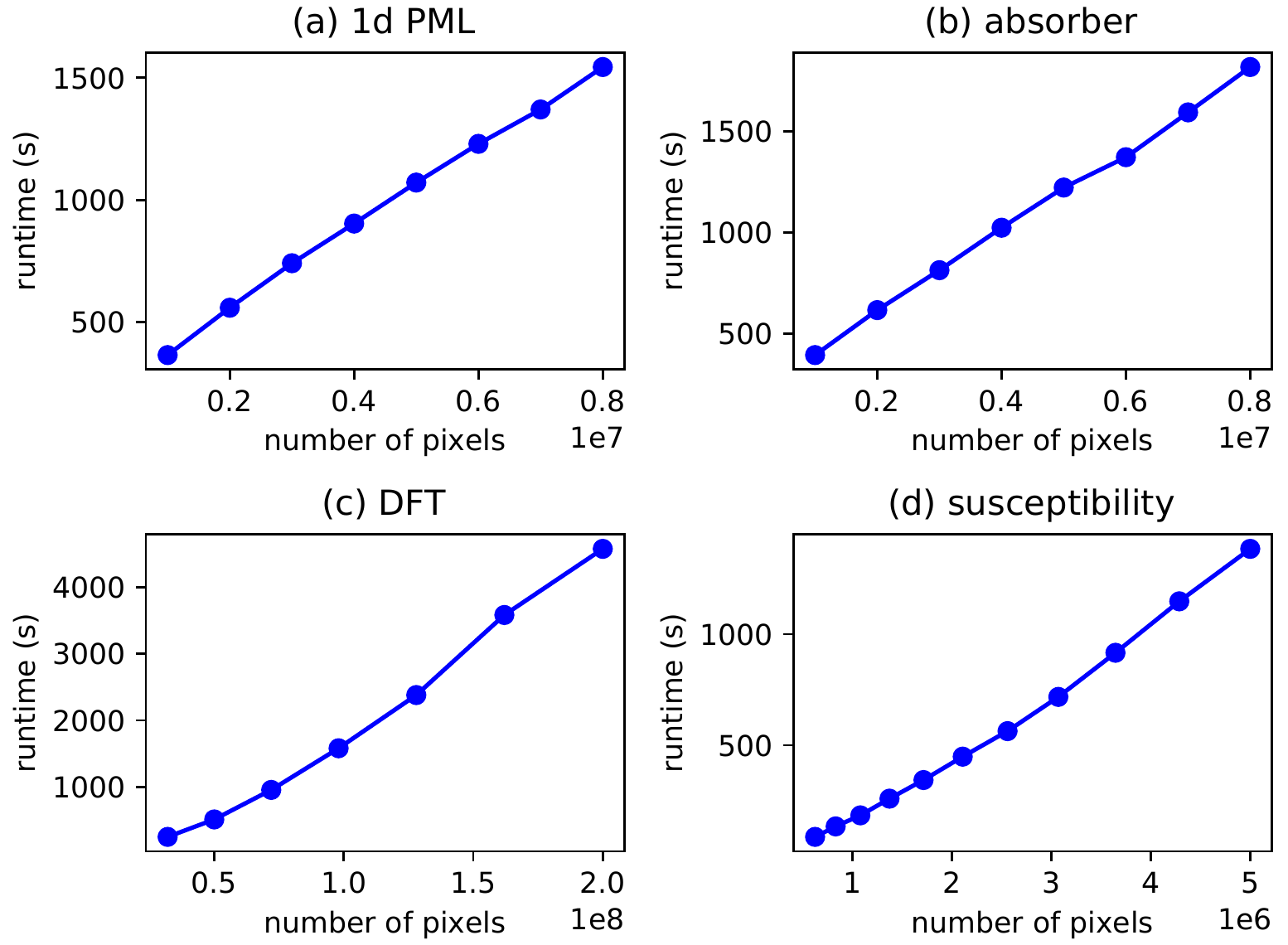}
\par}
\caption{Runtime (serial execution) vs. number of pixels of four single pixel-type 3d simulations: (a) 1d PML, (b) absorber (conductor), (c) discrete Fourier transform (DFT), and (d) susceptibility. The runtime scales nearly linearly with the number of pixels indicating the suitability of a linear-regression model for the load-balancing cost function of a given subvolume (\secref{LB}).
\label{fig:pixel-scaling}}
\end{figure}

\noindent The essential prerequisite of a load-balanced partitioning strategy is a way to estimate the computational work required by a given subvolume.  Given such an estimate, we can then employ the recursive-bisection method of \citeasnoun{Berger87}: recursively divide the domain in half along coordinate directions (typically along the current longest axis to minimize surface area, i.e. communication) so that the two halves contain equal work.  For non power-of-two numbers of processors $P$, we generalize this in \href{https://meep.readthedocs.io/en/latest/}{\emph{Meep}} to equal-work multisections (trisection etc.) according to the prime factors of $P$.  For finite-difference grids and other well-formed meshes, recursive bisection is provably within a small constant factor of the optimal surface area \cite{Simon97}.  In fact, we don't need to estimate the absolute computational work of a subvolume: we only need to estimate the \emph{relative} work of two subvolumes (the ratio). 

To estimate the relative costs of subvolumes, we constructed a linear regression model from a set of training data extracted from small serial executions. This model is based on the observation that the computations for different material features (e.g. nonlinearity or frequency dependence) and analyses (e.g. Fourier transforms) are performed separately for each grid point where they are needed in the FDTD algorithm, so their costs should be roughly additive and \emph{linear} in the number of affected grid points.  Under these assumptions, we could quickly extract regression coefficients by timing a set of eight 3d simulations for each parameter in which the number of grid points for only a single material was varied.   Four examples of these regression data are shown in \figref{pixel-scaling}; they exhibit the expected near-linear scaling.  Using this procedure, we extracted regression coefficients for 10 different features, representing the most expensive computations for typical simulations (including material anisotropy, nonlinearity, frequency-dependent materials, PML, Fourier analysis, and conductive materials).  The relative cost for any subvolume is then estimated by computing the number of pixels of each of these types and multiplying by the regression coefficients.

\begin{figure}[t]
{\centering \includegraphics[width=1.0\columnwidth]{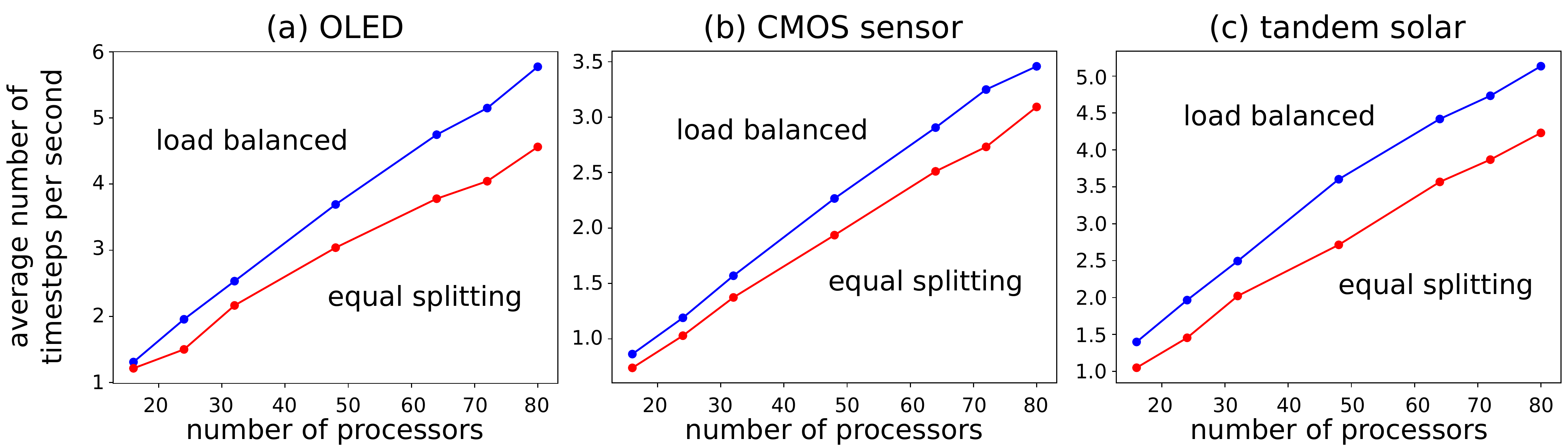}
\par}
\caption{Performance (timesteps/second) versus number of processors $P$ for three 3d physical simulations drawn from published works \cite{Oskooi2015,Yokogawa2017,Oskooi2014}. For each simulation, we consider both a naive partition into equal-volume chunks (``equal splitting'', red), and a load-balanced partition using a data-driven cost model (``load balanced'' splitting, blue). All cases exhibit about 30\% speedup due to load balancing (\secref{LB}).
\label{fig:speedup}}
\end{figure}

Given the cost estimate, we then recursively bisect the cell along one coordinate axis at a time to make the costs of the two halves as equal as possible (as determined by a binary search of cut positions, since the cost estimate is monotonic in the cut position).  We bisect the longest axis of the current domain in order to minimize surface area (communications). (As an exception, we bisect along a shorter axis if the load balancing is much better, by more than 30\%, along that axis.)

One can easily construct synthetic examples where load balancing improves performance by a factor of $P$, via problems that are essentially serialized by an equal partition.  For realistic computations, \emph{Meep} users have reported improvements by factors of~2 or more.  In \figref{speedup}, we show the improvements in three simulations drawn from real (published) applications, modeling an organic light-emitting diode (OLED)\cite{Oskooi2015}, a CMOS image sensor \cite{Yokogawa2017}, and a tandem solar cell \cite{Oskooi2014}.  All three simulations feature complicated materials in small regions of space and expensive Fourier (DFT) computations in other regions.  In all cases we obtain roughly linear scaling with the number of processors (using the Amazon EC2 clusters described in \secref{NN}, and roughly a 30\% improvement in performance compared to a naive equal partition.

Note also that we are exploiting the fact that, in electrodynamics simulations, the spatial arrangement of materials typically does not change in time. With the right data we can therefore ``statically'' load-balance the partition before the simulation begins. Other computational problems require more complicated dynamic load-balancing strategies, in which data is migrated between processors at runtime based on cost models or timing measurements \cite{Pearce12}. A static partition is much simpler and involved minimal modification to the \emph{Meep} codebase.

\section{Factorized Neural Nets for Performance Prediction}
\label{sec:NN}

\begin{figure}[t]
{\centering \includegraphics[width=0.9\columnwidth]{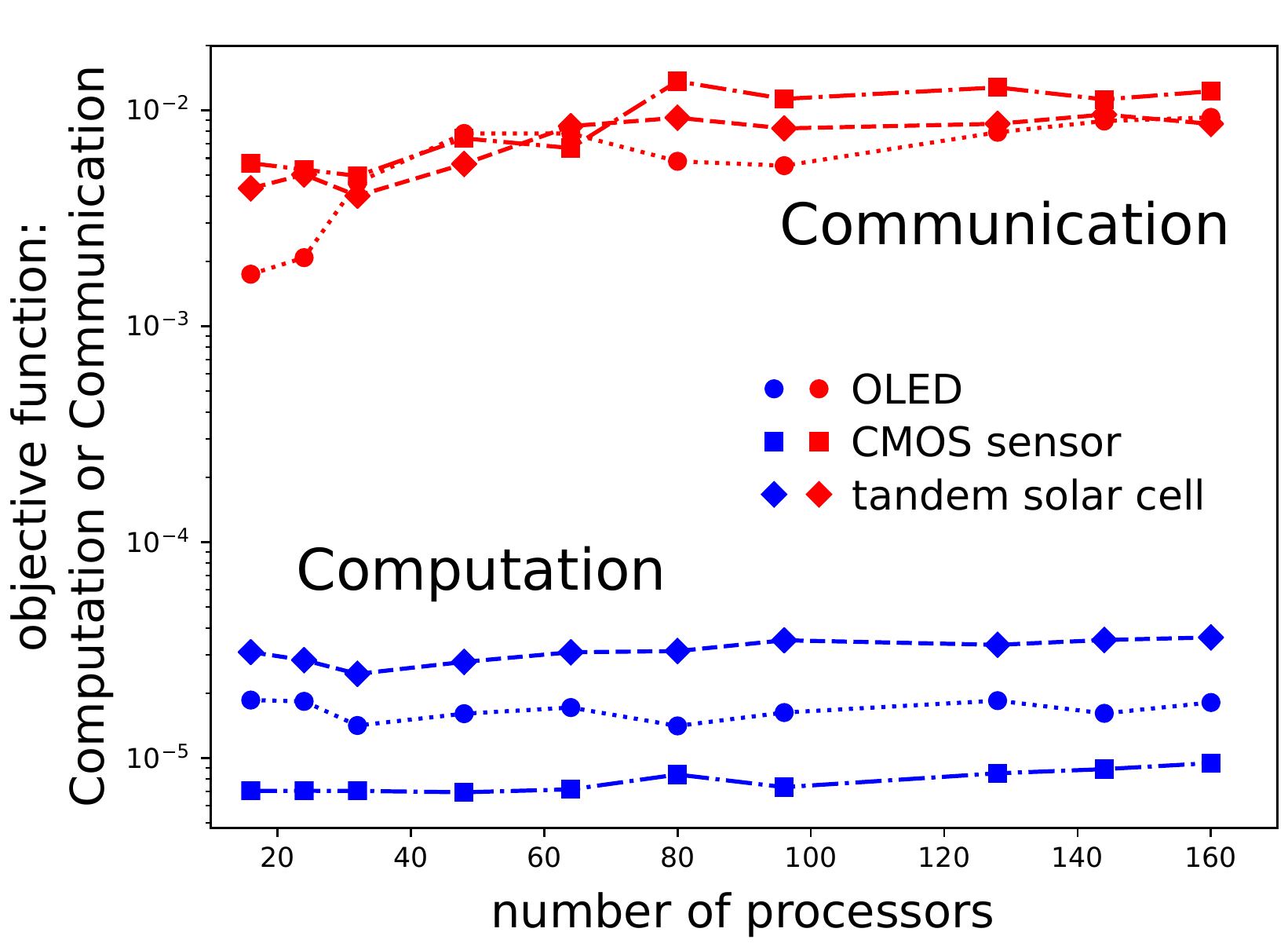}
\par}
\caption{Rescaled communication ($C$) and computation ($W$) time from \eqref{factorization} versus number of processors $P$ for three 3d physical simulations drawn from published works \cite{Oskooi2015,Oskooi2014,Yokogawa2017}. The fact that they are nearly independent of $P$ indicates that the asymptotic scalings factored out of \eqref{factorization} were effective; the remaining  variation is much easier for the NN to learn.
\label{fig:WC}}
\end{figure}

\noindent In this section, we describe the implementation, training, and validation of the NNs used for the $W(\vec{p})$ and $C(\vec{p})$ functions in the factorized execution time of \eqref{factorization}. The computation time is the sum of the time spent on (1) timestepping and (2) Fourier transforming the fields at selected points (for output analysis). The communication time is the time spent sending or receiving messages via MPI (or time spent waiting at synchronization barriers).   As was explained in \secref{intro}, the times were rescaled by rough asymptotic scaling factors to arrive at the coefficients $W$ and $C$ to be predicted by the NN. The expectation was that these coefficients would thereby have much less variation and hence be easier for a NN to interpolate from limited training data.  Evidence for this can be seen directly in \figref{WC}, where we show that $W$ and $C$ have relatively small dependence on the number $P$ of processors for several realistic problems. Note, however, that $W$ and $C$ still exhibit order-of-magnitude variations with \emph{other} parameters of the simulations $\vec{p}$, which the NNs must learn.   Below, we show that the resulting NNs can predict the execution time with reasonable accuracy; in contrast, we obtained errors many times larger when we initially attempted to train a \emph{single} NN to directly predict the total time $T(\vec{p})$.   At first, we also used a simple estimate $S \approx (N/P)^{2/3}$ of the average surface area, but we obtained a factor of $\sim 3$ improvement in $C$ prediction accuracy by scaling instead by the actual surface area computed by the load-balancing algorithm, which much more accurately reflects the influence of geometry on communication.

All \href{https://meep.readthedocs.io/en/latest/}{\emph{Meep}} simulations comprising the training and test data were benchmarked using Amazon Web Services (AWS) Elastic Compute Cloud (EC2) via MPICH clusters of c5.4xlarge instances (16 virtual CPUs, Intel Xeon Cascade Lake). \emph{AWS ParallelCluster} was used for the cluster setup and management. Each instance was running Amazon Linux with hyperthreading disabled (since \emph{Meep} is parallelized using MPI, there would be a substantial overhead to using more processes than physical cores). For all simulations, the computation and communication times were averaged over 100 timesteps. We verified the consistent performance of EC2 clusters by running three different test simulations 10 different times each. The coefficient of variation of the runtime was less than 1\% in all cases.

Rather than include all pixel types in the analysis, only a subset of the four most-common types found in physical simulations were used: susceptibility (for dispersive materials with complex, wavelength-dependent refractive index), discrete Fourier transform (for electromagnetic field monitors used to compute Poynting flux, near-to-far field transformations, energy density, etc.), PML with absorption in a single direction (for simulating open boundaries in the non-periodic direction of a cell with 2d periodic boundaries), and absorber boundary layers (for truncating dispersive materials and other cases in which PMLs often fail and require a workaround \cite{Loh2009}).  The total number $N$ of pixels and the number $P$ of processors were also NN inputs. Randomly generated simulations with random pixel-count distributions were used for training the NN. These random simulations are based on a 3d cell comprising three, non-overlapping, contiguous regions of: (1) crystalline silicon (susceptibility), (2) Poynting-flux monitor (DFT), and (3) isotropic permittivity with wavelength-independent refractive index of 3.5. The PML and absorber surround the cell in two different directions and overlap only vacuum. The training set consists of 6840 samples subdivided into 16, 24, 32, 48, and 64 processor-count samples (corresponding to 2, 3, 4, 6, and 8 instances) of 1346, 1338, 1798, 1637, and 721, respectively.

For validation, we used both additional random simulations as well as three physical simulations based on actual engineering applications: (1) visible-light extraction efficiency of an organic light-emitting diode (OLED) \cite{Oskooi2015}, (2) infrared-light absorption of a complimentary metal-oxide semiconductor (CMOS) image sensor \cite{Yokogawa2017}, and (3) light absorption of a tandem solar cell \cite{Oskooi2014}. The main variables in each of these physical simulations were the grid resolution, the number of frequencies of the DFT monitors, the size of the cell, and the thickness of the PML/absorber boundary layers.    Although the training was done using only random simulations, the distribution of pixel types in the random simulations was chosen to overlap the distributions found in the real engineering simulations to ensure that our NN was interpolating rather than extrapolating.  In addition, all of the training and test data has a communication/computation ratio $< 0.9$ to ensure that they are in the practically relevant regime of efficient parallelization.

\begin{figure}[t]
{\centering \includegraphics[width=1.0\columnwidth]{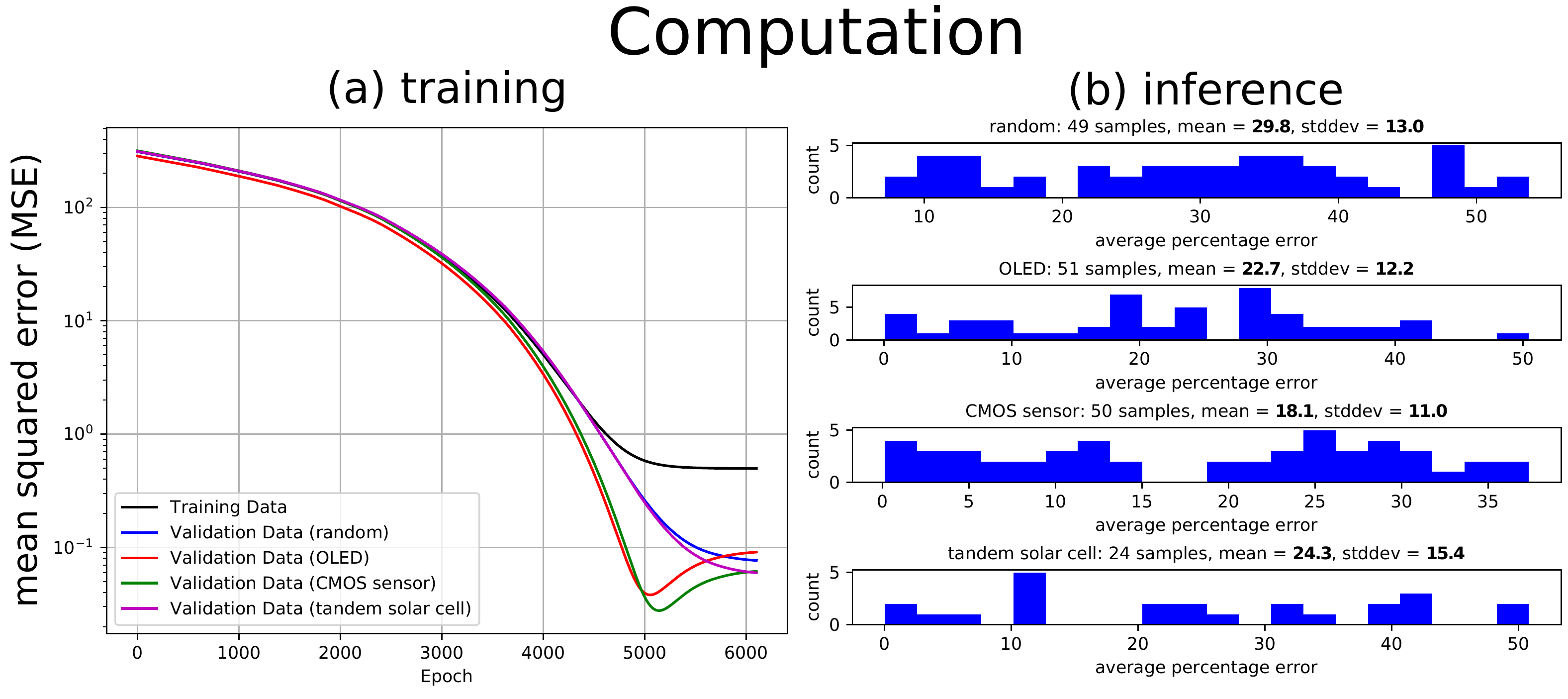}
\par}
\caption{Training and inference results of the optimal feed-forward neural network for the computation objective function $W(\vec{p})$ from \eqref{factorization}. (a) Evolution of the loss function [mean-squared error (MSE)] during training via back propagation. The training data (black) based on random simulations is shown along with four validation data. Increasing validation-data MSE and decreasing training-data MSE is an indication of overfitting. The training was stopped when the MSE of any one of the four validation data began to increase after reaching a minimum. (b) Histograms of the average percentage error of the inferred computation for four sets of test data consisting of random/non-physical and physical simulations.
\label{fig:W_training_inference}}
\end{figure}

\begin{figure}[t]
{\centering \includegraphics[width=1.0\columnwidth]{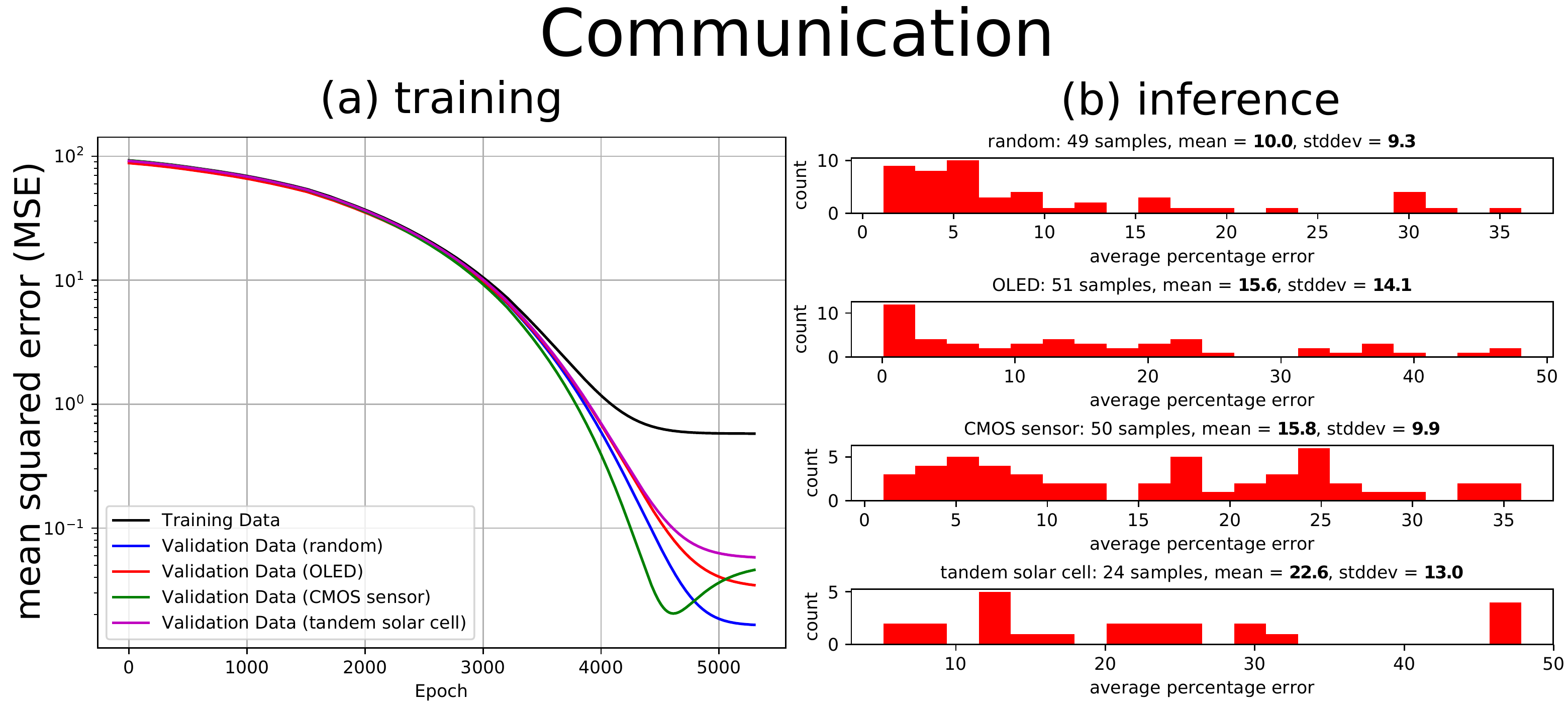}
\par}
\caption{Training and inference results of the optimal feed-forward neural network for the communication objective function $C(\vec{p})$ from \eqref{factorization}. (a) Evolution of the loss function [mean-squared error (MSE)] during training via back propagation. The training data (black) based on random simulations is shown along with four validation data. Increasing validation-data MSE and decreasing training-data MSE is an indication of overfitting. The training was stopped when the MSE of any one of the four validation data began to increase after reaching a minimum. (b) Histograms of the average percentage error of the inferred computation for four sets of test data consisting of random/non-physical and physical simulations.
\label{fig:C_training_inference}}
\end{figure}

\begin{figure}[t]
{\centering \includegraphics[width=0.9\columnwidth]{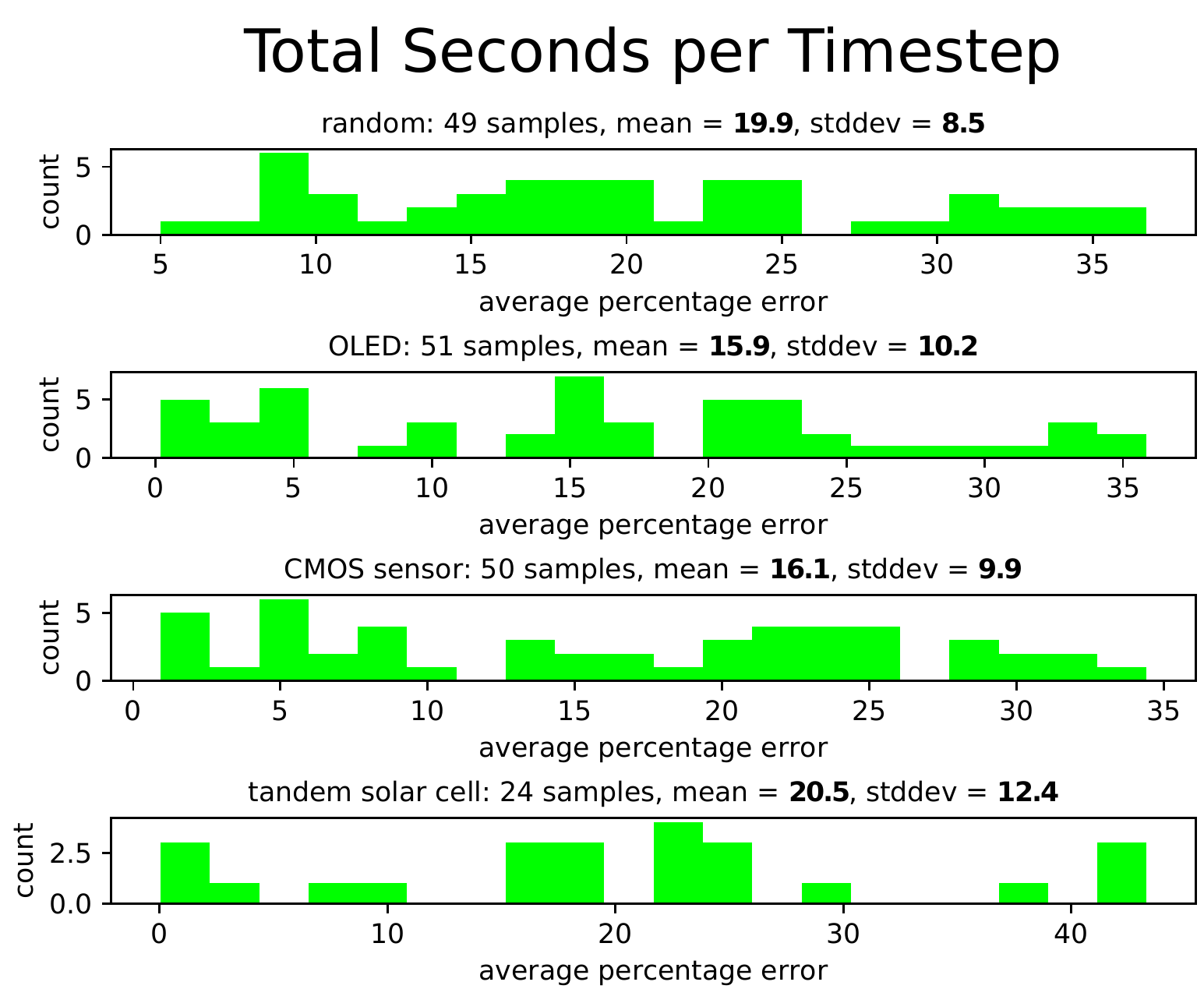}
\par}
\caption{Histograms of the average percentage error (APE) of the inferred total seconds per timestep for the four sets of test data obtained by combining the results from the computation and communication neural networks. The APE is $\approx 20\pm10\%$ for all test data.
\label{fig:avg_sec_per_step_error}}
\end{figure}

The machine-learning model was based on feed-forward NNs implemented via the \emph{PyTorch} framework \cite{paszke2017automatic}, with an activation function of a leaky rectified linear unit (ReLU), using a loss function of a mean-squared error (MSE) in $\log W$ or $\log C$ as explained in \secref{intro}. There are six inputs to each NN, grouped into two categories: (1) the fractional number of the four pixel types (normalized by the total number of pixels in the cell), (2) the total number of pixels, and (3) the number of processors. The inputs are also normalized using a log transform, both to limit the range of the inputs and also to allow the network to construct power-law models via linear combinations. The optimal NN architecture for the computation and communication functions was determined by sweeping over the number of layers, neurons per layer, learning rate, and momentum. The training was stopped when the MSE of any one of the four validation data began to increase which is an indication of overfitting. \Figref{W_training_inference} shows the training and inference results for the optimal NN for the computation function which consists of two hidden layers with 30 and 60 neurons in each layer. The Adam optimizer is used with a learning rate of 0.000012 and momentum of 0.9. The number of epochs is 6100. \Figref{C_training_inference} shows training and inference results for the optimal NN for the communication function which consists of three hidden layers with 20, 30, and 40 neurons in each layer. The Adam optimizer is used with a learning rate of 0.00001 and momentum of 0.9. The number of epochs is 5300. We also tried other optimizers such as RMSProp, AdaGrad, etc. but these were not found to be optimal. Training and inference results for the computation and communication NNs are shown in \figreftwo{W_training_inference}{C_training_inference}. \Figref{avg_sec_per_step_error} shows the error histogram for the inferred total seconds per timestep obtained by combining the results for the computation and communication NNs.

The final result in \figref{avg_sec_per_step_error} is that we are always able to predict the total execution time $T$ within a factor of two, with a typical mean ($\approx$~median) error of around $20\%$, even for realistic simulations that had no direct analogue in the random training data.  For comparison, rather than train two separate NNs for the computation and communication, we also trained a single NN whose output was simply $T(\vec{p})$, the execution time per timestep, using the same training data. The results were found to be considerably worse, with a mean percentage error well above 100\% in most cases.  The largest errors are for the computation $W$ and we discuss in \secref{conclusions} how the $W$ accuracy could be improved. Conversely, if we omit the $N$ parameter from the NNs, we found that we could obtain nearly undiminished prediction accuracy.  Although including more inputs arguably improves NN generality, there might be some advantage to employing only the fraction of each computation type: scale-invariant fractions allow simulations of very different sizes to interpolate within the same parameter space.

\section{Concluding Remarks}
\label{sec:conclusions}

\noindent There are many possible avenues to refine the approach described in this work. The simplest would be to use more accurate scale factors in \eqref{factorization}. For example, instead of scaling the work by the number $N$ of grid points we could use the linear-regression estimate from the load-balancing procedure (\secref{LB}), which scales different types of grid points by different coefficients.  Incorporating more accurate scaling into the execution-time formulas should make training the neural network easier and more accurate.  Similar to ``analytical'' performance models~\cite{Sanjay08}, one could use a more fine-grained factorization of the execution time, with individual NNs to predict different portions of the computation or communication, at the cost of greater code instrumentation. That being said, predicting performance to within $\approx 20\pm10\%$ is generally accurate enough for the purpose of efficiently allocating computational resources and choosing simulation parameters.

Conversely, one could imagine using the improved performance estimate provided by the neural network, which incorporates communications costs, to further improve the load-balancing procedure (for which we currently only use a linear model of computational cost).  That idea, however, leads to a feedback loop: changing the load balancing procedure means changing the execution time, which requires re-training the network, which then alters load balancing again, and so forth, greatly increasing the amount of training data required. The current regression-based load-balancing procedure seems to be a good balance between complexity and performance in \emph{Meep}.  More generally, other scientific simulation codes may have very different performance characteristics from the one considered here, but we believe that the key lesson of incorporating asymptotic scaling knowledge will still be pertinent. One almost always knows roughly how the time scales (linearly, inversely, etc.) with different types of simulation parameters, and a neural network is an effective ``black box'' into which all the \emph{remaining} complications can be stuffed.

Although we demonstrated this approach with a specific (albeit practically important) type of scientific computation, we believe that a similar approach is applicable to many other computational modeling tasks. Mathematically, \emph{Meep}'s FDTD timesteps have the same structure as a sparse matrix--vector multiplication~\cite{oskooi2010meep} arising from a mesh with local interactions; this is the dominant computation in a vast array of scientific applications, such as iterative linear solvers for finite-element models. Predicting the performance of many FDTD ``materials'' is thus equivalent to modeling sparse matrices with a variety of heterogeneous sparsity structures. More generally, in almost every large-scale computational problem one knows the asymptotic scaling of the computation and communication costs as a function of the number of processors and the size of the data, and our key point is that one should factor out these crude scalings before applying machine-learning techniques to performance predictions.

\section*{Acknowledgments}
\noindent This work was made possible by the National Science Foundation (NSF) via Small Business Innovation Research (SBIR) Phase I and II awards 1647206 and 1758596, and contributes to the PAPPA program of DARPA MTO under award HR0011-20-90016. This work was also supported by the AWS Cloud Credits for Research Program.

\end{document}